\documentclass[aps,prd,twocolumn,nofootinbib,showpacs]{revtex4-1}

\usepackage{graphicx,epsfig,color,amsmath}

\usepackage[colorlinks=true,linkcolor=blue,citecolor=blue, urlcolor=blue]{hyperref}

\begin{document}

\title{Total decay width of $H \to gg$ using the infinite-order scale-setting approach based on the intrinsic conformality}

\author{Chu-Tian Gao$^1$}
\email{gaoct@cqu.edu.cn}

\author{Xing-Gang Wu$^{1}$}
\email{wuxg@cqu.edu.cn}

\author{Xu-Dong Huang$^1$}
\email{hxud@cqu.edu.cn}

\author{Jun Zeng$^2$}
\email{zengj@cqu.edu.cn}

\affiliation{$^1$ Department of Physics, Chongqing Key Laboratory for Strongly Coupled Physics, Chongqing University, Chongqing 401331, People's Republic of China}
\affiliation{$^2$ INPAC, Key Laboratory for Particle Astrophysics and Cosmology (MOE), Shanghai Key Laboratory for Particle Physics and Cosmology, School of Physics and Astronomy, Shanghai Jiao Tong University, Shanghai 200240, People's Republic of China}

\date{\today}

\begin{abstract}

We make a detailed study on the properties of the total decay width of Higgs decay channel $H\to gg$ up to $\alpha_s^6$-order QCD corrections by using the newly suggested infinite-order scale-setting approach, which is based on the ideas of both the principle of maximum conformality and the intrinsic conformality. This approach is called as the PMC$_\infty$ approach. By using the PMC$_\infty$ approach, we observe that the conventional renormalization scale ambiguity in perturbative QCD calculation is eliminated and the residual scale dependence due to unknown higher-order terms can also be highly suppressed. We then obtain an accurate perturbative QCD prediction on the total decay width, e.g. $\Gamma (H \to gg)|_{\rm PMC_\infty} =336.42^{+7.01}_{-6.92}$ KeV, where the errors are squared averages of those from all the mentioned error sources.

\end{abstract}

\maketitle

In quantum chromodynamics (QCD), the Higgs boson plays an important role in precision test of the Standard Model (SM), and it is also helpful for searching the new physics beyond the SM. The Higgs boson decays into two gluons is an important channel for studying the Higgs phenomenology~\cite{Djouadi:2005gi}. The coupling of
the Higgs to gluons is predominantly mediated by the top quark within the SM, and the high-order QCD corrections to this process can be evaluated in an effective theory in which the top quark has been integrated out~\cite{Inami:1982xt}. At present, the perturbative QCD (pQCD) correction to the total decay width of the Higgs decay channel $H \to gg$, e.g. $\Gamma(H \to gg)$, has been calculated up to next-to-next-to-next-to-next-to-leading order (${ \rm N^{4}LO}$) in the limit of a large top-quark mass~\cite{Djouadi:1991tka, Graudenz:1992pv, Dawson:1993qf, Spira:1995rr, Dawson:1991au, Chetyrkin:1997iv, Chetyrkin:1997un, Baikov:2006ch, Herzog:2017dtz}. We are then facing the opportunity of achieving precise pQCD prediction on $\Gamma(H \to gg)$.

It is helpful to reduce the pQCD uncertainties as much as possible. Among them, the error caused by using conventional scale-setting approach is usually treated as an important systematic error for pQCD prediction. Such error in making fixed-order prediction occurs because one conventionally assumes an arbitrary renormalization scale to do the numerical analysis, which is usually chosen as the typical momentum flow of the process or the one assumed to be the effective virtuality of the strong interaction or the one to eliminate large logs so as to achieve a more convergent series, and etc.. This {\it ad hoc} assignment of renormalization scale causes the mismatching of the $\alpha_s$ and the corresponding coefficients, thus the coefficients of the QCD running coupling at each order strongly depend on the choice of renormalization scale as well as the renormalization scheme. However, as indicated by the renormalization group invariance, a physical observable must be independent to the choice of renormalization scale. In the literature, the principle of maximum conformality (PMC)~\cite{Brodsky:2011ta, Brodsky:2012rj, Mojaza:2012mf, Brodsky:2013vpa} has been suggested to remove such renormalization scale ambiguity. It is well-known that the $\alpha_s$-running behavior is governed by the renormalization group equation (RGE). Then the $\{\beta_i\}$-terms emerged in the pQCD series can be inversely adopted for fixing the correct $\alpha_s$-value of a high-energy process. The purpose of PMC is to rightly determine the effective coupling constant of the process (whose argument is called as the PMC scale) with the help of RGE~\cite{Wu:2013ei, Wu:2014iba}, whose prediction is found to be independent to any choice of renormalization scale and satisfies the requirement of renormalization group invariance. The PMC scale-setting procedure agrees with the standard scale-setting procedure of Gell-Mann and Low~\cite{Gell-Mann:1954yli} in the QED Abelian limit (small number of colors, $N_C \to 0$~\cite{Brodsky:1997jk}).

Many successful PMC applications have been done in the literature. Previously, the PMC has been applied for dealing with the decay width $\Gamma(H\to gg)$~\cite{Wang:2013bla, Zeng:2015gha, Zeng:2018jzf}. It is noted that the PMC was originally introduced as a multi-scale approach, in which distinct effective couplings (and hence the PMC scales) at each order have been derived due to different categories of $\{\beta_i\}$-terms occur at each order. Furthermore, because the same category of $\{\beta_i\}$-terms emerges at different orders, the determined PMC scales are in perturbative form. This leads to the fact that the precision of the PMC scales at higher orders decrease with the increment of perturbative orders, since fewer $\{\beta_i\}$-terms are known for fixing the value of higher-order $\alpha_s$. Thus the PMC multi-scale approach shall have explicit residual scale dependence~\cite{Zheng:2013uja}, and if the convergence of the perturbative series of the PMC scale is weak, such residual scale dependence could be large~\cite{Wu:2019mky}.

By further taking the intrinsic conformality (iCF) property into PMC, a new infinite-order scale-setting approach, called as the PMC$_\infty$ approach, has been recently proposed in the literature~\cite{DiGiustino:2020fbk}. The PMC$_\infty$ approach follows from the PMC, its resultant conformal coefficients are the same as the PMC ones at each perturbative order, but sets the effective PMC scales at each order by requiring all the scale-dependent $\{\beta_i\}$-terms at each order to vanish exactly and separately~\cite{DiGiustino:2020fbk}. Via this way, the newly fixed PMC scales at each order are in definite form and are no-longer in perturbative series, thus the residual scale dependence of the previous PMC scales due to their previous perturbative nature can be exactly eliminated. This indicates that the precision of the previous PMC predictions on the total decay width $\Gamma(H\to gg)$~\cite{Wang:2013bla, Zeng:2015gha, Zeng:2018jzf} may be further improved by applying the PMC$_\infty$ approach. It is thus interesting to make a detailed study on $\Gamma(H\to gg)$ by using the PMC$_\infty$ approach.

Practically, the decay width of the Higgs decays into two gluons at the $\alpha_s^6$-order level can be expressed as
\begin{equation}
\Gamma  (H\to gg)=\frac{M_H^3 G_F}{36 \sqrt{2} \pi }\left[\sum _{k=0}^{4 }  C_{k}(\mu_r) a^{k+2}_{s}(\mu_r)\right],   \label{rij}
\end{equation}
where Fermi constant $G_F=1.16638 \times 10^{-5}~{\rm{GeV}}^{-2}$, $a_s=\alpha_s/4\pi$, and $\mu_r$ stands for an arbitrary renormalization scale. The perturbative coefficients $C_{k\in[0,4]}(\mu_r)$ at the initial scale of $\mu_r=M_H$ under conventional $\overline{\rm MS}$-scheme can be read from Refs.\cite{Inami:1982xt, Djouadi:1991tka, Graudenz:1992pv, Dawson:1993qf, Spira:1995rr, Dawson:1991au, Chetyrkin:1997iv, Chetyrkin:1997un, Baikov:2006ch, Herzog:2017dtz}. As has been argued in Refs.\cite{Wang:2013bla, Zeng:2015gha, Zeng:2018jzf}, it is important to firstly transform them into the ones under a physical momentum space subtraction scheme (mMOM-scheme)~\cite{Celmaster:1979km, Celmaster:1979dm, Celmaster:1979xr, Celmaster:1980ji, Gracey:2013sca, vonSmekal:2009ae} such that to avoid the ambiguities of fixing the PMC scales with the help of RGE. The mMOM-scheme is gauge dependent, a detailed discussion of gauge dependence after applying the PMC has been done in Ref.\cite{Zeng:2020lwi}, which shows that if the gauge parameter $\xi\in[-1,1]$, the mMOM prediction shall have weaker $\xi$-dependence. And for definiteness, we adopt the Landau gauge ($\xi=0$) to do the analysis, whose corresponding coefficients $C_{k}(\mu_r)$ at any renormalization scale $\mu_r$ can be achieved by recursively applying the RGE. The explicit expressions for the required coefficients up to $\alpha_s^6$-order level can be found in Refs.\cite{Zeng:2015gha, Zeng:2018jzf}.

Due to the iCF property, we can divide the ${\rm N^4LO}$-level total decay width into five conformal subsets,
\begin{equation}
\Gamma (H\to gg)=\frac{M_H^3 G_F}{36 \sqrt{2} \pi }\sum\limits_{n={\rm I}}^{\rm V}\Gamma_{n},
\end{equation}
which collect together the same category of non-conformal terms into each subset and ensure the scheme independence of each subset via the commensurate scale relations among different orders~\cite{Brodsky:1994eh}. Each conformal subset satisfies the scale invariant condition,
\begin{eqnarray}
\left(\mu^2_r \frac{ \partial}{\partial \mu^2_r} +\beta(\alpha_s)\frac{\partial}{\partial \alpha_s}\right) \Gamma_n=0.
\label{sigmainvariance}
\end{eqnarray}
More explicitly, we have
\begin{widetext}
\begin{eqnarray}
\Gamma_{\rm I}&=&A_{{\rm Conf}}\bigg[a_{s}^2(\mu_r)+2 B_{\beta_0} \beta _0 a_{s}^3(\mu_r)+\big(3 B_{\beta_0}^2 \beta_0^2 +2 B_{\beta_0} \beta _1 \big)a_{s}^4(\mu_r)+ \big(7 B_{\beta_0}^2 \beta _1 \beta _0 +4 B_{\beta_0}^3 \beta_0^3  \nonumber \\
&&\quad\quad\quad +2 B_{\beta_0} \beta _2\big) a_{s}^5(\mu_r) +\big(8 B_{\beta_0}^2 \beta _2  \beta _0 +\frac{47}{3} B_{\beta_0}^3 \beta_1  \beta _0^2 +5 B_{\beta_0}^4 \beta_0^4 +2 B_{\beta_0} \beta _3 +4 B_{\beta_0}^2 \beta_1^2\big) a_{s}^6(\mu_r)\bigg],   \\
\Gamma_{\rm II}&=&B_{{\rm Conf}}\bigg[a_{s}^3(\mu_r)+3 C_{\beta_0} \beta _0 a_{s}^4(\mu_r)+\big(6 C_{\beta_0}^2 \beta_0^2 +3 C_{\beta_0} \beta _1 \big)a_{s}^5(\mu_r)+ \bigg(\frac{27}{2} C_{\beta_0}^2 \beta _1  \beta _0 + 10 C_{\beta_0}^3 \beta_0^3 +3 C_{\beta_0} \beta_2\bigg) a_{s}^6(\mu_r)\bigg],  \\
\Gamma_{\rm III}&=&C_{{\rm Conf}}\bigg[a_{s}^4(\mu_r)+4 D_{\beta_0} \beta _0 a_{s}^5(\mu_r) + \big(10 D_{\beta_0}^2\beta_0^2 +4 D_{\beta_0} \beta_1 \big)a_{s}^6(\mu_r)\bigg],  \\
\Gamma_{\rm IV}&=&D_{{\rm Conf}}\bigg[a_{s}^5(\mu_r)+5 E_{\beta_0}\beta _0 a_{s}^6(\mu_r)\bigg],  \\
\Gamma_{V}&=&E_{{\rm Conf}}\bigg[a_{s}^6(\mu_r)\bigg].
\end{eqnarray}
\end{widetext}
Here $A_{{\rm Conf}}$, $B_{{\rm Conf}}$, $C_{{\rm Conf}}$, $D_{{\rm Conf}}$ and $E_{{\rm Conf}}$ are conformal coefficients, and $B_{\beta_0}=\ln{{\mu_r^2}/{\mu_{\rm I}^2}}$, $C_{\beta_0}=\ln{{\mu_r^2}/{\mu_{\rm II}^2}}$, $D_{\beta_0}=\ln{{\mu_r^2}/{\mu_{\rm III}^2}}$, $E_{\beta_0}=\ln{{\mu_r^2}/{\mu_{\rm IV}^2}}$. The PMC$_\infty$ scales $\mu_{\rm I,\cdots,IV}$, which can be fixed by using the scale invariant condition (\ref{sigmainvariance}). To match the mMOM-scheme perturbative series, the $\{\beta_i\}$-functions under the mMOM-scheme should be adopted, whose explicit forms up to five-loop level are available in Ref.\cite{Ruijl:2017eht}. Then, following the standard PMC$_\infty$ scale-setting procedures, the conformal coefficients and the PMC$_\infty$ scales can be derived from the known coefficients $C_{k}$ via a step-by-step manner. For examples, we have $A_{{\rm Conf}}=C_{0}$; The conformal coefficient $B_{{\rm Conf}}$ can be determined by setting $n_f=\frac{33}{2}$\footnote{Due to the $\beta_0=11-\frac{2}{3}n_f$, to remove the $\beta_0$-dependent terms from the coefficients $C_i$ is equivalent to set $n_f=\frac{33}{2}$. } to drop off the $\beta_0$ terms in $C_{1}$, and the PMC$_\infty$ scale $\mu_{\rm I}$ can be fixed by using the known conformal coefficients $A_{{\rm Conf}}$, $B_{{\rm Conf}}$ and the $\{\beta_0\}$-terms of $C_{1}$; and etc.. For convenience, we put all the required conformal coefficients and the PMC$_\infty$ scales in the Appendix.

Then, we can transform the original perturbative series (\ref{rij}) into the following conformal series
\begin{widetext}
\begin{eqnarray}
\Gamma &&(H \to gg)=\frac{M_H^3 G_F}{36 \sqrt{2} \pi }\bigg[A_{{\rm Conf}}a_{s}^2(\mu_{\rm I})+B_{{\rm Conf}}a_{s}^3(\mu_{\rm II}) + C_{{\rm Conf}}a_{s}^4(\mu_{\rm III})+D_{{\rm Conf}}a_{s}^5(\mu_{\rm IV})+E_{{\rm Conf}}a_{s}^6(\mu_{\rm V})\bigg].
\end{eqnarray}
\end{widetext}
The PMC$_\infty$ scales are definite and have no perturbative nature, thus they exactly avoid the residual scale ambiguity due to unknown higher-order terms in the perturbative series of the original PMC scales. Numerically, the first four PMC$_\infty$ scales are
\begin{eqnarray}
\{\mu_{\rm I}, \mu_{\rm II}, \mu_{\rm III}, \mu_{\rm IV} \} = \{50.1, 46.0, 63.0, 61.3\} (\rm GeV),
\end{eqnarray}
which are invariant to any choice of renormalization scale and avoid conventional renormalization scale ambiguity. It is interesting to note that those PMC$_\infty$ scales are around $M_H \exp(-5/6)\sim 54$ GeV, which is suggested by the Gell-Mann Low scheme~\cite{Gell-Mann:1954yli}, in which $\exp(-5/6)$ is a result of the convention that has been chosen to define the minimal dimensional regularization scheme. At present, the PMC$_\infty$ scale $\mu_{\rm V}$ at the highest order can not be determined, since there is no $\{\beta_i\}$-terms to fix its magnitude. As usual, we adopt $\mu_{\rm V}=\mu_{\rm IV}$~\cite{Brodsky:2013vpa}, which ensures the scheme independence of the resultant conformal series. Numerically, we have found that due to the coefficient $E_{{\rm Conf}}$ is free of divergent renormalon terms, the magnitude of the final term is negligibly small, and the uncertainty of the total decay width caused by different choice of $\mu_{\rm V}$ is negligible.

To do the numerical calculation, we take the top-quark pole mass $M_t = 172.5 \pm0.7~\rm{GeV}$, and $M_H = 125.25 \pm 0.17$~$\rm{GeV}$~\cite{Workman:2022zbs}. The QCD asymptotic scale $\Lambda$ can be determined by using the world average of $\alpha_s$ at the scale $M_Z$, e.g. $\alpha_s(M_Z)=0.1179\pm 0.0009$~\cite{Workman:2022zbs}. As a subtle point, we need to transform the asymptotic scale from the $\overline{\rm MS}$-scheme to the mMOM-scheme by using the Celmaster-Gonsalves relation~\cite{Celmaster:1979km, Celmaster:1979dm, Celmaster:1979xr, Celmaster:1980ji}.

\begin{widetext}
\begin{figure}[htb]
\centering
\includegraphics[width=0.450\textwidth]{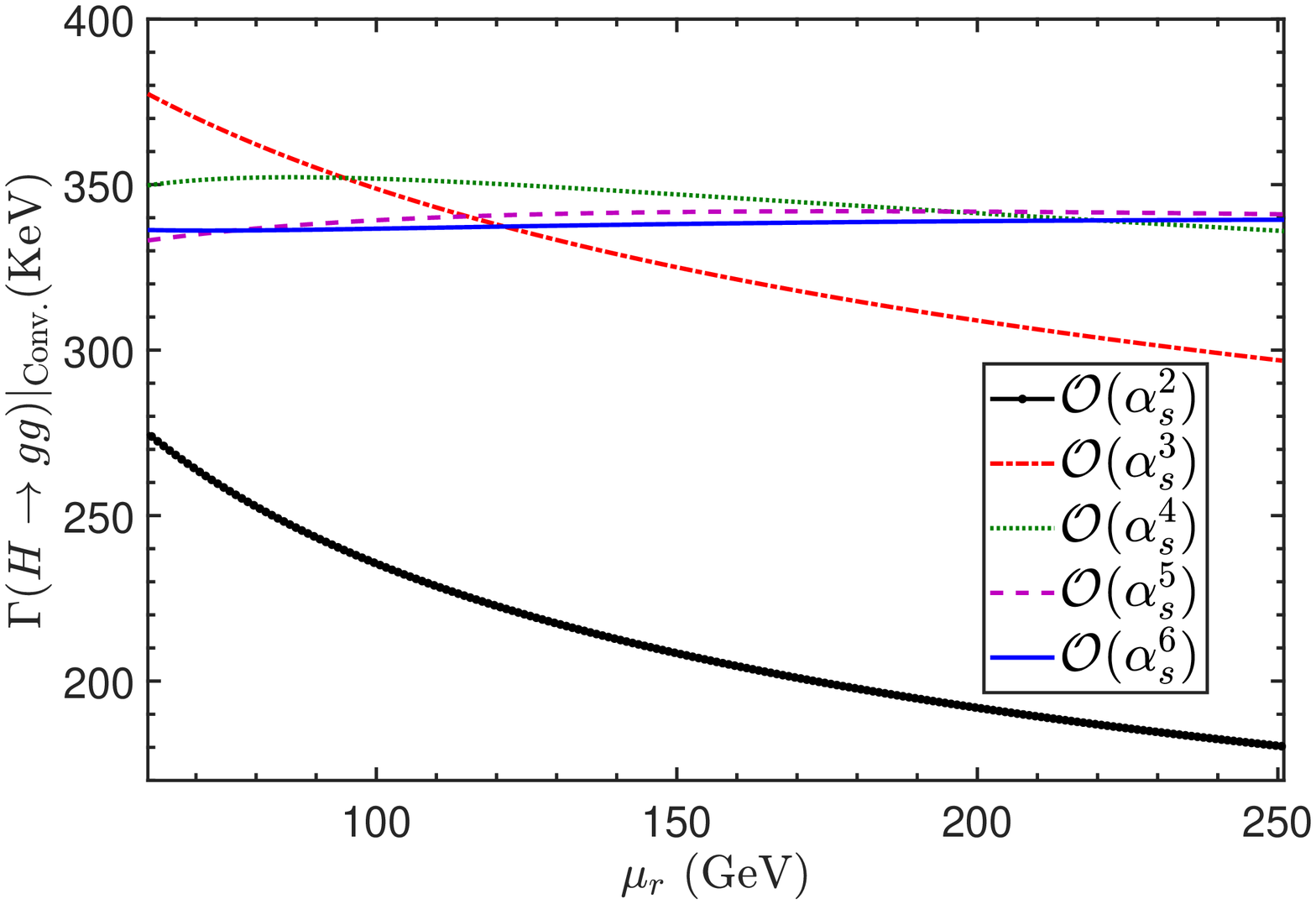}
\includegraphics[width=0.450\textwidth]{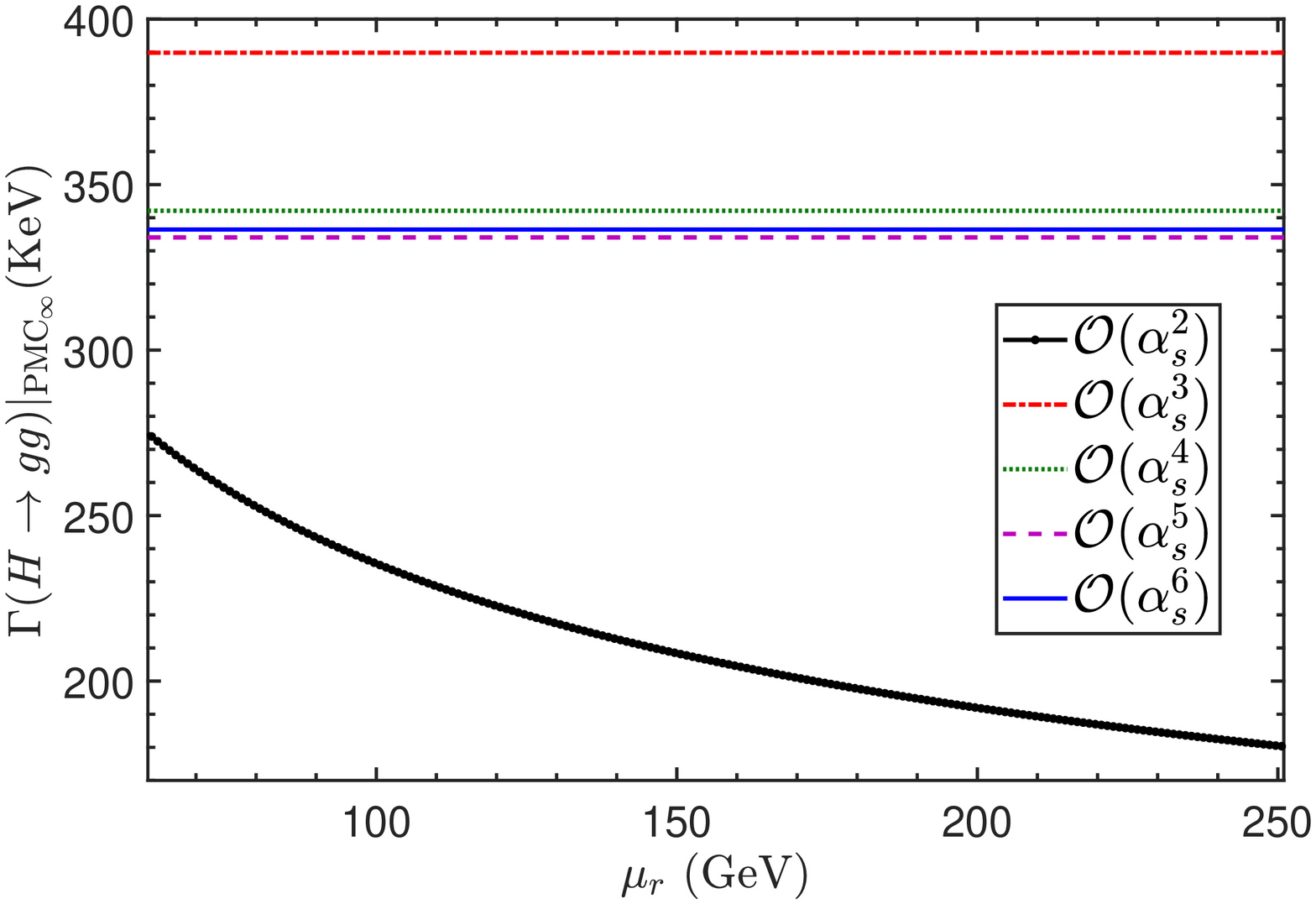}
\caption{The decay width $\Gamma (H \to gg)$ under conventional and PMC scale-setting approaches, respectively. The solid line with big dot, the dash-dot line, the dotted line, the dashed line and the solid line are predictions up to $\mathcal{O}(\alpha_s^2)$, $\mathcal{O}(\alpha_s^3)$, $\mathcal{O}(\alpha_s^4)$, $\mathcal{O}(\alpha_s^5)$, and $\mathcal{O}(\alpha_s^6)$, respectively.}
\label{HtoggConv}
\end{figure}
\end{widetext}

By setting all input parameters to be their central values, we firstly present the decay width $\Gamma (H \to gg)$ up to different $\alpha_s$-orders under conventional (Conv.) and PMC$_{\infty}$ scale-setting approaches in Fig.~\ref{HtoggConv}. At the $\mathcal{O}(\alpha_s^2)$-order level, the perturbative series of $\Gamma (H \to gg)$ does not have $\{\beta_i\}$-terms to fix $\mu_{\rm I}$, and the PMC$_\infty$ and conventional predictions are the same and both of them are scale dependent. Fig.~\ref{HtoggConv} shows that the decay width $\Gamma (H \to gg)$ under conventional scale-setting approach has a strong dependence on $\mu_r$, which becomes smaller and smaller, when more and more loop terms have been included. Fig.~\ref{HtoggConv} also shows that the decay width $\Gamma (H \to gg)$ at $\mathcal{O}(\alpha_s^3)$-order and higher orders under PMC$_\infty$ scale-setting shall be independent to any choice of renormalization scale, due to the fact that the scale-dependent noconformal terms have been exactly eliminated.

\begin{widetext}
\begin{center}
\begin{table}[htb]
\begin{tabular}{cccccccccc}
\hline
& $n=2$ & $n=3$ & $n=4$ & $n=5$ & $n=6$ & $\kappa_1$  & $\kappa_2$ & $\kappa_3$ & $\kappa_4$ \\
 \hline
  $\Gamma^{\mathcal{O}(\alpha_s^{n})}|_{\rm Conv.}$ & $219.86^{+54.05}_{-39.50}$ & $335.46^{+41.34}_{-38.59}$ & $349.71^{+2.52}_{-13.68}$ & $340.95^{+1.00}_{-7.67}$ & $337.45^{+1.94}_{-1.18}$  & $[38\%,65\%]$ & $[0, 13\%]$ & $[0, 4.8\%]$ & $[0, 1.0\%]$   \\
 \hline
  $\Gamma^{\mathcal{O}(\alpha_s^{n})}|_{\rm PMC_\infty}$ & $219.86^{+54.05}_{-39.50}$ & $389.86$ & $342.09$ & $334.05$ & $336.42$  & $[53\%, 95\%]$ & $12\%$ & $2.4\%$ & $0.7\%$ \\
 \hline
\end{tabular}
\caption{Results for the decay width $\Gamma (H \to gg)$ (in unit: KeV) and $\kappa_n$ up to different loop corrections under conventional and PMC$_\infty$ scale-setting approaches, respectively. The NLO and higher order PMC$_\infty$ predictions are scale independent; While under conventional scale-setting approach, the central values are for $\mu_r=M_H$, and the errors are for $\mu_r\in[M_H/2, 2M_H]$. }
\label{results}
\end{table}
\end{center}
\end{widetext}

\begin{figure}[htb]
\centering
\includegraphics[width=0.50\textwidth]{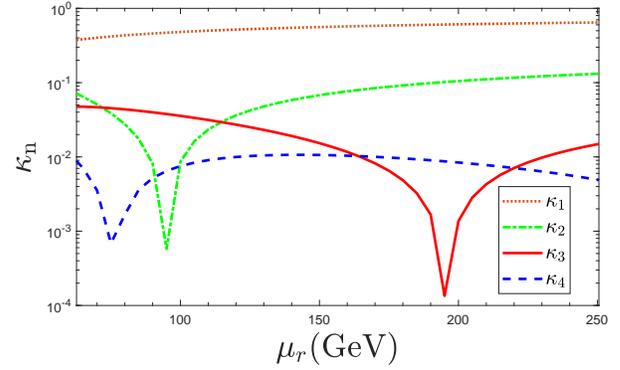}
\caption{Results for the ratio $\kappa_n$ versus the renormalization scale $\mu_r$ under conventional scale-setting approach up to different loop corrections. The ratios under PMC$_\infty$ approach are scale invariant. }
\label{kappaC}
\end{figure}

Secondly, we present the decay width $\Gamma (H \to gg)$ up to different loop QCD corrections under conventional and PMC$_\infty$ scale-setting approaches in Table~\ref{results}. To show the perturbative property, we define a ratio
\begin{eqnarray}
\kappa_{n} = \left|\frac{\Gamma^{\mathcal{O}(\alpha_s^{n+2})}- \Gamma^{\mathcal{O}(\alpha_s^{n+1})}}{\Gamma^{\mathcal{O}(\alpha_s^{n+1})}}\right| ,
\end{eqnarray}
which indicates how the ``known" prediction $\Gamma^{\mathcal{O}(\alpha_s^{n+1})}$ is affected by the one-order-higher terms. As for the PMC$_\infty$ series, we have $\kappa_1>\kappa_2>\kappa_3>\kappa_4$ for any choice of $\mu_r$, indicating the relative difference between the two nearby orders becomes smaller when more loop terms have been included. This feature is consistent with the perturbative nature of the series and indicates that one can obtain more precise prediction by including more loop terms. As for the conventional series, as shown by Fig.~\ref{kappaC}, there are crossovers for $\kappa_{2,3,4}$ within the range of $\mu_r\in[M_H/2, 2M_H]$, and the ratios vary from $0$ to $13\%$, $4.8\%$ and $1.0\%$ for $\kappa_2$, $\kappa_3$ and $\kappa_4$, respectively.

\begin{widetext}
\begin{center}
\begin{table}[htb]
\begin{tabular}{ccccccc}
\hline
  & ~~~~$\rm LO$~~~~ & ~~~~$\rm NLO$~~~~ & ~~~~$\rm N^2LO$~~~~ & ~~~~$\rm N^3LO$~~~~ & ~~~~$\rm N^4LO$~~~~ & ~~~~$\rm Total$~~~~ \\
  \hline
  ~~$\Gamma^{\mathcal{O}(\alpha_s^{6})}|_{\rm Conv.}$~~ & $218.66^{-40.95}_{+57.39}$ & $116.62^{+2.08}_{-15.59}$ & $14.44^{+25.23}_{-41.60}$ & $-8.70^{+13.84}_{-8.02}$ & $-3.57^{+1.74}_{+6.64}$ & $337.45^{+1.94}_{-1.18}$  \\
  \hline
  ~~$\Gamma^{\mathcal{O}(\alpha_s^{6})}|_{\rm PMC_\infty}$~~ &  299.57  &  95.22  &  $-$41.82  &  $-$19.59  &  3.04  &  336.42  \\
  \hline
  ~~$\Gamma^{\mathcal{O}(\alpha_s^{6})}|_{\rm PMCm}$~~ &  289.83$^{+0.96}_{-0.24}$  &  91.19$^{-0.51}_{+0.29}$  &  $-$31.78$^{+0.30}_{-0.11}$  &  $-$13.36  &  1.92  &  337.80$^{+0.75}_{-0.06}$ \\
  \hline
\end{tabular}
\caption{The values (in unit: KeV) of each loop-term (LO, NLO, N$^2$LO and N$^3$LO) for the four-loop prediction $\Gamma^{\mathcal{O}(\alpha_s^{6})}$ under the conventional, PMC$_\infty$ scale-setting approaches and PMCm scale-setting approaches, respectively. The PMC$_\infty$ predictions are scale independent; While under conventional and PMCm scale-setting approaches, the central values are for $\mu_r=M_H$, and the errors are for $\mu_r\in[M_H/2, 2M_H]$. }
\label{resultsorder}
\end{table}
\end{center}
\end{widetext}

Moreover, to show the convergence of the perturbative series explicitly, we present the magnitudes of each loop terms for the four-loop approximants $\Gamma^{\mathcal{O}(\alpha_s^{6})}$ in Table~\ref{resultsorder}. Table~\ref{resultsorder} shows that the relative importance of the LO-terms: NLO-terms: N$^2$LO-terms : N$^3$LO-terms : N$^4$LO-terms for conventional series are
\begin{displaymath}
 1 : +53.3^{+13.5}_{-16.7}\% : +6.6^{+15.7}_{-16.4}\% : -4.0^{+6.9}_{-2.1}\% : -1.6^{+0.6}_{-2.7}\%,
\end{displaymath}
where the central values are for $\mu_r=M_H$, and the errors are for $\mu_r\in [\frac{1}{2}M_H, 2M_H]$. The scale dependence for each loop terms are large, but due to the cancellation of scale dependence among different orders, the net scale dependence is small, e.g. $(^{+0.6\%}_{-0.3\%})$ for $\mu_r\in [\frac{1}{2}M_H, 2M_H]$. On the other hand, there are no renormalization scale dependence for each loop terms of the PMC$_{\infty}$ prediction $\Gamma^{\mathcal{O}(\alpha_s^{6})}$. More explicitly, we have also presented the values of each loop-terms (LO, NLO, N$^2$LO, N$^3$LO or N$^4$LO) for $\Gamma^{\mathcal{O}(\alpha_s^{6})}$ under the PMC$_{\infty}$ approach in Table~\ref{resultsorder}. At the four-loop level, the PMC$_\infty$ series already represents good convergent behavior, and the relative importance of the LO-terms: NLO-terms: N$^2$LO-terms : N$^3$LO-terms : N$^4$LO-terms becomes
\begin{displaymath}
 1 : +31.8\%: -14.0\%: -6.5\%: +1.0\%,
\end{displaymath}
whose magnitudes are scale invariant, indicating that the PMC$_\infty$ perturbative series represents the intrinsic perturbative behavior of $\Gamma (H \to gg)$. As a comparison, we also show the numerical result under the PMC multi-scale approach (PMCm) in Table~\ref{resultsorder}, which still has some residual scale-dependence. However, its numerical effects is smaller than the conventional one. Detailed formulas for the PMCm approach can be found in Ref.\cite{Zeng:2018jzf}.

Thirdly, after eliminating the renormalization scale ambiguities, there are still some other error sources for the pQCD prediction of the total decay width, such as the $\alpha_s$ fixed-point error $\Delta\alpha_s(M_Z)$, the Higgs mass uncertainty $\Delta M_{H}$, the top-quark pole mass uncertainty $\Delta M_t$, and etc. Up to $\alpha_s^6$-order, we have
\begin{eqnarray}
\Gamma(H\to gg)|_{\rm Conv.}&=& 337.45 ^{+6.27+1.21+0.02}_{-6.20-1.21-0.02} \;\;{\rm KeV}, \\
\Gamma(H\to gg)|_{\rm PMC\infty} &=& 336.42^{+6.21+1.22+0.03}_{-6.14-1.20-0.01} \;\;{\rm KeV},
\end{eqnarray}
where the errors are for $\Delta \alpha_s(M_Z) =\pm 0.0009$ (which leads to $\Lambda^{n_f=5}_{\rm mMOM}=362.0^{+36.6}_{-18.0}$ MeV), $\Delta M_{H}=\pm0.17$ GeV, and $\Delta M_t=\pm0.7$ GeV, respectively. Here the conventional predictions are achieved by fixing $\mu_r=M_H$.

Using the PMC$_\infty$ approach, the PMC scales at each orders are no longer evaluated as a perturbative series, thus avoiding the \textit{first kind of residual scale dependence}. As has been mentioned above, all the PMC$_\infty$ scales are around $M_H \exp(-5/6)\sim 54$ GeV, so the \textit{second kind of residual scale dependence} is small due to the convergent behavior at higher orders. As a further step of making a conservative estimation on the contributions from the uncalculated ${n+1}_{\rm th}$-order terms, we take its PMC$_\infty$ scale $\mu_{n}$ to be within the region of the latest determined PMC$_\infty$ scale $\mu_{n-1}$, e.g. $\mu_{n}\in[\mu_{n-1}/2, 2\mu_{n-1}]$ and take $\Delta\Gamma^{{\mathcal O}(\alpha_s^{i+1})}|_{\rm PMC_\infty}=\pm\frac{M_H^3 G_F}{36\sqrt{2}\pi}|C_{i, {\rm Conf}}a^{i+2}_s(\mu_{i+1})|_{\rm MAX}$ with $i=I, II, III, IV, V$, respectively. Numerically, we obtain $\Delta\Gamma^{{\mathcal O}(\alpha_s^{2})}|_{\rm PMC_\infty}=\pm 2.99$ KeV, $\Delta\Gamma^{{\mathcal O}(\alpha_s^{3})}|_{\rm PMC_\infty}=\pm 1.80$ KeV, $\Delta\Gamma^{{\mathcal O}(\alpha_s^{4})}|_{\rm PMC_\infty}=\pm 1.25$ KeV, $\Delta\Gamma^{{\mathcal O}(\alpha_s^{5})}|_{\rm PMC_\infty}=\pm 0.47$ KeV, $\Delta\Gamma^{{\mathcal O}(\alpha_s^{6})}|_{\rm PMC_\infty}=\pm 0.09$ KeV. It can be found that the estimated errors may underestimate the contributions listed in Table~\ref{results}, which need to be multiplied by $\{57, 27, 6, 5\}$, respectively. Thus, the numerical result of $\Delta\Gamma^{{\mathcal O}(\alpha_s^{6})}|_{\rm PMC_\infty}$ need to be multiplied by $5$ as analogy. Similarly, one can obtain the corresponding values for conventional scale-setting approach, which are $\pm 11.98$ KeV, $\pm 4.60$ KeV, $\pm 1.40$ KeV, $\pm 0.73$ KeV and $\pm 0.14$ KeV, respectively. In order to match with the center values shown in Table~\ref{results}, they need to be multiplied by $\{10, 3, 6, 5\}$. And the numerical result of $\Delta\Gamma^{{\mathcal O}(\alpha_s^{6})}|_{\rm Conv.}$ need to be multiplied by $5$ as analogy. Those values are slightly larger than the PMC$_\infty$ ones due to larger perturbative coefficient $C_k$ than the conform coefficient at each orders even though it is compensated by a smaller $\alpha_s$ value.

As a summary, we have presented a detailed analysis of the Higgs-boson decay $H \to gg$ up to $\alpha_s^6$-order, and we obtain
\begin{eqnarray}
\Gamma (H \to gg)|_{\rm Conv.}&=& 337.45^{+6.67}_{-6.43}\;\; {\rm KeV}, \\
\Gamma (H \to gg)|_{\rm PMC_\infty}  &=& 336.42^{+6.33}_{-6.26} \;\;{\rm KeV},
\end{eqnarray}
where the errors are squared averages of those from $\Delta \alpha_s(M_Z)$, $\Delta M_{H}$, $\Delta M_t$ and the uncertainty of the renormalization scale within the region of $\left[{M_H}/{2},2 M_H\right]$. The errors are dominated by $\Delta \alpha_s(M_Z)$, then followed by the choice of renormalization scale and the accuracy of Higgs mass. If the value of $\alpha_s(M_Z)$ can be measured accurately to avoid the error from $\Delta \alpha_s(M_Z)$, we will obtain
\begin{eqnarray}
\Gamma (H \to gg)|_{\rm Conv.}&=& 337.45^{+2.29}_{-1.69}\;\; {\rm KeV}, \\
\Gamma (H \to gg)|_{\rm PMC_\infty}  &=& 336.42^{+1.22}_{-1.20} \;\;{\rm KeV}.
\end{eqnarray}

The Higgs-boson decay $H \to gg$ provides another successful example for the application of PMC$_\infty$ scale-setting method to high-energy processes. Up to N$^4$LO QCD corrections, the pQCD predictions under the PMC$_\infty$ and conventional scale-setting approaches are consistent with each other. But the conventional renormalization scale uncertainties are still sizable, which are about $1\%$ by varying the renormalization scale $\mu_r$ within the range of $[\frac{1}{2}M_H, 2M_H]$. By applying the PMC$_\infty$, the $\alpha_s$ values at lower orders are definitely fixed by the requirement of intrinsic conformality, the conventional renormalization scale ambiguity is eliminated, and the residual scale dependence from the original PMC multi-scale-setting approach is also highly suppressed. Thus a more precise test of the SM can be achieved.

\hspace{0.5cm}

\noindent {\bf Acknowledgments:} This work was supported by the Chongqing Graduate Research and Innovation Foundation under Grant No.CYB21045 and No.ydstd1912, by the Natural Science Foundation of China under Grant No.12175025 and No.12147102, by the Fundamental Research Funds for the Central Universities under Grant No.2020CQJQY-Z003 and No.2021CDJZYJH-003. \\

\appendix

\section*{Appendix: The conformal coefficients and PMC$_\infty$ scales up to $\alpha_s^6$-order level}

Applying the PMC$_\infty$ scale-setting approach together with the general ``degeneracy" pattern of the QCD theory~\cite{Bi:2015wea}, the perturbative series of the decay width $\Gamma(H\to gg)$ under mMOM-scheme is
\begin{widetext}
\begin{eqnarray}
\Gamma (H\to gg)&=&\frac{M_H^3 G_F}{36 \sqrt{2} \pi }\bigg[ A_{{\rm Conf}}a_{s}^2(\mu_r)+\left(B_{{\rm Conf}}+2 A_{{\rm Conf}}B_{\beta_0} \beta _0 \right) a_{s}^3(\mu_r) +\bigg(C_{{\rm Conf}}+3 B_{{\rm Conf}}C_{\beta_0} \beta _0 +3 A_{{\rm Conf}}B_{\beta_0}^2 \beta_0^2\nonumber \\
&&+2 A_{{\rm Conf}}B_{\beta_0} \beta _1\bigg) a_{s}^4(\mu_r) +\bigg(D_{{\rm Conf}}+7 A_{{\rm Conf}}B_{\beta_0}^2 \beta _1 \beta _0 +4 C_{{\rm Conf}}D_{\beta_0} \beta _0+6 B_{{\rm Conf}}C_{\beta_0}^2 \beta_0^2+4 A_{{\rm Conf}}B_{\beta_0}^3 \beta_0^3\nonumber \\
&& +2 A_{{\rm Conf}}B_{\beta_0} \beta _2 +3 B_{{\rm Conf}}C_{\beta_0} \beta _1\bigg) a_{s}^5(\mu_r)+\bigg(E_{{\rm Conf}}+8 A_{{\rm Conf}}B_{\beta_0}^2 \beta _2  \beta _0 +\frac{27}{2} B_{{\rm Conf}}C_{\beta_0}^2 \beta _1  \beta _0 +5 D_{{\rm Conf}} E_{\beta_0}\beta _0  \nonumber\\
&&+\frac{47}{3} A_{{\rm Conf}}B_{\beta_0}^3 \beta_1  \beta _0^2+10 C_{{\rm Conf}}D_{\beta_0}^2\beta_0^2 +10 B_{{\rm Conf}}C_{\beta_0}^3 \beta_0^3+5 A_{{\rm Conf}}B_{\beta_0}^4 \beta_0^4 +2 A_{{\rm Conf}}B_{\beta_0} \beta _3 +3 B_{{\rm Conf}}C_{\beta_0} \beta_2  \nonumber\\
&&+4 A_{{\rm Conf}}B_{\beta_0}^2 \beta_1^2 +4 C_{{\rm Conf}}D_{\beta_0} \beta_1\bigg) a_{s}^6(\mu_r)\bigg]+\mathcal{O}[a_{s}^7(\mu_r)], \label{betarij}
\end{eqnarray}
\end{widetext}

To compare Eq.~(\ref{rij}) with Eq.~(\ref{betarij}), one can determine the conformal coefficients and PMC$_\infty$ scales for $\Gamma(H\to gg)$ up to $\alpha_s^6$-order level via a step-by-step manner, i.e.
\begin{widetext}
\begin{eqnarray}
&&A_{{\rm Conf}}=C_0, \\
&&B_{{\rm Conf}}=C_1\left(n_f=\frac{33}{2}\right), \\
&&C_{{\rm Conf}}=C_2\left(n_f=\frac{33}{2}\right)-2 A_{{\rm Conf}}B_{\beta_0}\bar{\beta _1} , \\
&&D_{{\rm Conf}}=C_3\left(n_f=\frac{33}{2}\right)-2 A_{{\rm Conf}}B_{\beta_0} \bar{\beta _2}-3 B_{{\rm Conf}}C_{\beta_0}\bar{\beta _1} , \\
&&E_{\rm Conf} = C_4\left(n_f=\frac{33}{2}\right) -2 A_{\rm Conf} B_{\beta_0} \bar\beta_3 - 3B_{\rm Conf} C_{\beta_0} \bar\beta_2 - 4A_{\rm Conf} B_{\beta_0}^2 \bar\beta_1^2-4 C_{\rm Conf} D_{\beta_0} \bar\beta_1
\end{eqnarray}
and
\begin{eqnarray}
&&\ln{\frac{\mu_r^2}{\mu_{\rm I}^2}}=\frac{C_1-B_{{\rm Conf}}}{2 A_{{\rm Conf}} \beta_0} , \\
&&\ln{\frac{\mu_r^2}{\mu_{\rm II}^2}}=\frac{C_2-C_{{\rm Conf}}-3 A_{{\rm Conf}}B_{\beta_0}^2 \beta_0^2-2 A_{{\rm Conf}}B_{\beta_0} \beta _1}{3 B_{{\rm Conf}} \beta _0} ,  \\
&&\ln{\frac{\mu_r^2}{\mu_{\rm III}^2}}=\frac{C_3-D_{{\rm Conf}}-7 A_{{\rm Conf}}B_{\beta_0}^2 \beta _1 \beta _0-6 B_{{\rm Conf}}C_{\beta_0}^2 \beta_0^2-4 A_{{\rm Conf}}B_{\beta_0}^3 \beta_0^3 -2 A_{{\rm Conf}}B_{\beta_0} \beta _2 -3 B_{{\rm Conf}}C_{\beta_0} \beta _1}{4 C_{{\rm Conf}}\beta _0} ,  \\
&&\ln{\frac{\mu_r^2}{\mu_{\rm IV}^2}}=(C_4-E_{{\rm Conf}}-8 A_{{\rm Conf}}B_{\beta_0}^2 \beta _2 \beta _0 -\frac{27}{2} B_{{\rm Conf}}C_{\beta_0}^2 \beta _1 \beta _0 -\frac{47}{3} A_{{\rm Conf}}B_{\beta_0}^3 \beta_1 \beta _0^2-10 C_{{\rm Conf}}D_{\beta_0}^2 \beta_0^2-10 B_{{\rm Conf}}C_{\beta_0}^3 \beta_0^3
\nonumber  \\
&&-5 A_{{\rm Conf}}B_{\beta_0}^4 \beta_0^4 -2 A_{{\rm Conf}}B_{\beta_0} \beta _3 -3 B_{{\rm Conf}}C_{\beta_0} \beta_2
-4 A_{{\rm Conf}}B_{\beta_0}^2 \beta_1^2 -4 C_{{\rm Conf}}D_{\beta_0} \beta_1 )/{5 D_{{\rm Conf}}\beta _0 }   ,
\end{eqnarray}
\end{widetext}
Here $\bar{\beta _1}=\beta _1(n_f=\frac{33}{2})=-107$, $\bar{\beta _2}=\beta _2(n_f=\frac{33}{2})=-2001.29$ and $C_{k}$ are the perturbative coefficients.

\end{document}